\documentclass[aps,reprint,prd,twocolumn,floatfix,nofootinbib]{revtex4}%

\usepackage{amsfonts}
\usepackage{amsmath}
\usepackage{amssymb}
\usepackage{graphicx}
\usepackage[ansinew]{inputenc}
\usepackage{pdfpages}
\usepackage{epsfig}
\usepackage{subfig}
\RequirePackage{filecontents}

\usepackage[colorlinks,hyperindex]{hyperref}%

\providecommand{\U}[1]{\protect\rule{.1in}{.1in}}

\hypersetup{
colorlinks,
citecolor=blue,
linkcolor=black,
urlcolor=black,
}

\begin{document}

\title{Study of the charged super-Chandrasekhar limiting mass white dwarfs in the $f(R,\mathcal{T})$ gravity}

\author{F. Rocha$^{\star}$, G. A. Carvalho$^{\ddagger}$ $^{\star}$, D. Deb$^{\dagger,a}$ and M. Malheiro$^{\star}$}
\affiliation{$^\star$ITA - Instituto Tecnol\'ogico de Aeron\'autica, 12228-900, S\~ao Jos\'e dos Campos, SP, Brazil\\ $^\ddagger$Instituto de Pesquisa e Desenvolvimento (IP\&D), Universidade do Vale do Para\'iba, 12244-000, S\~ao Jos\'e dos Campos, SP, Brazil\\ $^\dagger$Department of Physics, Indian Institute of Engineering Science 
	and Technology, Shibpur, Howrah 711103, West Bengal, India}

\email{$^a$ddeb.rs2016@physics.iiests.ac.in}


\begin{abstract}
The equilibrium configuration of white dwarfs composed of a charged perfect fluid are investigated in the context of the $f(R,\mathcal{T})$ gravity, for which $R$ and $\mathcal{T}$ stand for the Ricci scalar and the trace of the energy-momentum tensor, respectively. By considering the functional form $f(R, \mathcal{T})=R+2\chi \mathcal{T}$, where $\chi$ is the matter-geometry coupling constant, and for a Gaussian ansatz for the electric distribution, some physical properties of charged white dwarfs were derived, namely: mass, radius, charge, electric field, effective pressure and energy density; their dependence on the parameter $\chi$ was also derived. In particular, the $\chi$ value important for the equilibrium configurations of charged white dwarfs has the same scale of $10^{-4}$ of that for non-charged stars and the order of the charge was $10^{20}$C, which is scales with the value of one solar mass, i.e., $\sqrt{G}M_\odot\sim 10^{20}$C. We have also showed that charged white dwarf stars in the context of the $f(R,\mathcal{T})$ have surface electric fields generally below the Schwinger limit of $1.3\times 10^{18}$V/m. In particular, a striking feature of the coupling between the effects of charge and $f(R,\mathcal{T})$ gravity theory is that the modifications in the background gravity increase the stellar radius, which in turn diminishes the surface electric field, thus enhancing stellar stability of charged stars. Most importantly, our study reveals that the present $f(R,T)$ gravity model can suitably explain the super-Chandrasekhar limiting mass white dwarfs, which are suppose to be the reason behind the over-luminous SNeIa and remain mostly unexplained in the background of general relativity (GR).
 
\end{abstract}
\maketitle


\section{Introduction}\label{sec:int}

With the recent pioneering observations, such as supernovae of type Ia~\cite{Perlmutter1999,Bennett2003}, baryon acoustic oscillations~\cite{Percival2010}, Planck data~\cite{Ade2014}, Cosmic Microwave Background Radiation (CMBR)~\cite{Spergel2003,Spergel2007} and redshift supernovae~\cite{Riess1998} it is evident that presently our Universe is going through the accelerated expanding phase which hardly can be explained through the most successful theory of GR. The most standard way out to explain the present observed cosmological dynamics appeared as the inclusion of the cosmological constant $(\Lambda)$ into the Einstein gravitational field equation which also provided fine agreement with the observed data by considering the presence of a hypothetical component known as \emph{Dark matter}~\cite{Overduin2004,Baer2015}. It is also largely accepted that the sole reason behind the present accelerated expansion phase of the Universe is actually another mysterious component widely known as Dark Energy~\cite{Sahni2004,Bamba2012,Copland2006,Amendola2010,Friemen2008} and appeared as the most successful avenue in explaining the present cosmic dynamical phase until it faced the major set back due to a huge mismatch of the values of 120 orders of magnitude between the observationally achieved and theoretically predicted values of $\Lambda$~\cite{Weinberg1989,Carroll2001}.

To overcome this situation different researchers came up with more sophisticated gravity theories by modifying the Einstein-Hilbert action which gave rise a new avenue known as modified/extended gravity theories. The extended theories of gravity have aroused as an opportunity to solve problems which are still without convincing explanation within GR framework. The most famous modified theory of gravity is the $f(R)$ theory, which consists of choosing a more general action to replace the Einstein-Hilbert one, this is made by assuming that the gravitational action is given by an arbitrary function of the Ricci scalar $R$ can be found in literature Refs.~\cite{Capozziello2002,Nojiri2003,Carroll2004, Bertolami2007}. Besides $f(R)$ gravity theory in recent times the extended gravity models attracted the attention of the researchers are $f(R,G)$ gravity~\cite{Nojiri2005,Bamba2010}, Brans-Dicke (BD) gravity~\cite{Avilez2014,Bhattacharya2015}, $f(\mathbb{T})$ gravity~\cite{Bengochea2009,Linder2010,Bohmer2011}, etc., where $G$ and  $f(\mathbb{T})$ are Gauss-Bonnet and torsion scalar, respectively.

Recently, Harko et al. \cite{Harko2011} developed a further generalization of the $f(R)$ theory of gravity by choosing a gravitational action as an arbitrary function of the Ricci scalar and also the trace of the energy-momentum tensor $\mathcal{T}$, which is called $f(R,\mathcal{T})$ theory of gravity. Within this theory, Solar System tests have been already performed \cite{Deng2015,Shabani2014}. Studies on compact astrophysical objects have also been considered in the literature \cite{Moraes2016,Carvalho2017,Deb2018Dec,Deb2019Mar}. In particular, modified theories of gravity have been shown to significantly elevate the maximum mass of compact objects \cite{Capozziello2016Jan,Carvalho2019Mar,Das2015May,Deb2019Mar}, which means that $f(R,\mathcal{T})$ is of particular interest for the hydrostatic equilibrium configuration of compact stars.

In what concerns to white dwarfs they are the final evolution state of main sequence stars with initial masses up to $8.5-10.6 M_{\odot}$. However, if the WD mass grows over 1.44 M$_{\odot}$ - known as Chandrasekhar mass limit \cite{Chandrasekhar1931} - as in binary systems, where the main star is receiving mass from a nearby star, a type Ia
supernova (SNIa) explosion may occur. However, with the recently observed peculiar highly over-luminous SNeIa, such as, SN 2003fg, SN 2006gz, SN 2007if, SN 2009dc~\cite{Howell2006,Scalzo2010} it is possible to confirm the existence of a huge Ni-mass which leads to the possibility of massive super-Chandrasekhar white dwarfs with mass $2.1-2.8~M_\odot$ as their most feasible progenitors.

To provide some physical mechanism where a super-Chandrasekhar white dwarf could support the gravitational collapse a lot of works have bubbled in the literature with different proposals. To cite some of them, we have: general relativistic \cite{Carvalho2018Apr,Boshkayev2014Sep}, strong magnetic field \cite{Das2013Feb,Das2012Aug,Franzon2015Oct,Franzon2017Feb,Chatterjee2017Mar,Otoniel2019Jul,Bera2016Jan}, modified theories of gravity \cite{EslamPanah2019May,Liu2019Feb,Carvalho2017Dec,Das2015May,Banerjee2017Oct,Kalita2018Sep}, background gravity corrections \cite{Ray2019Feb}, rotation, differential rotation and charge effects \cite{Liu2014,Carvalho2018}. 

In addition, several authors have studied charged stars. Within them, there are investigations about the influence of the electrical charge distribution at the stellar structure of polytropic stars \cite{Ray2003, Arbanil2015, Azam2016}, anisotropic stars \cite{Deb2018epjc}, strange stars \cite{Arbanil2015, Malheiro2011} and white dwarfs \cite{Liu2014,Carvalho2018}. In what concerns to charged WDs, Liu and collaborators \cite{Liu2014} found that the charge contained in WDs can affect their structure, they have larger masses and radii than the uncharged ones. Moreover, Carvalho et al. have shown in a previous work \cite{Carvalho2018} that the increment of the total charge from $0$ to $\approx 2\times10^{20}$C allows to increase the total mass in approximately $55.58\%$, and for a large total charge, more massive stellar objects are found.

Some works have also approached the coupling between charge and $f(R,\mathcal{T})$ gravity effects for stellar equilibrium \cite{Sharif2017,Deb2018Dec,Abbas2019Apr,Sharif2018Sep}. Those works showed in particular, that charged objects have more stable configurations than non-charged ones. They also showed that the energy conditions are respected inside the compact objects.

Here in this work, we are particularly interested to study the charge effects within the framework of the $f(R,\mathcal{T})$ gravity, for the hydrostatic equilibrium configurations of white dwarfs. To do so, we revise the formalism of the $f(R,\mathcal{T})$ gravity in section \ref{secII}, showing the basic equations and deriving the hydrostatic equilibrium configurations for the charged case. In section \ref{secIII} we describe stellar properties that we assume, namely, equation of state and electric charge distribution. In section \ref{secIV} we outline our results and in section \ref{secV} we present our conclusions.  

\section{Basic formalism}\label{secII}
\subsection{$f(R,\mathcal{T})$ gravity}
The modified form of the Einstein-Hilbert action in the Einstein-Maxwell space-time is as follows:
\begin{multline}\label{1.1}
S=\frac{1}{16\pi}\int d^{4}xf(R,\mathcal{T})\sqrt{-g}+\\ \int d^{4}x\mathcal{L}_m\sqrt{-g}+\int d^{4}x\mathcal{L}_e\sqrt{-g},
\end{multline}
where $T_{\mu\nu}$ is the energy-momentum tensor of the matter distribution, ${\mathcal{L}}_m$ represents the Lagrangian for the matter distribution and ${\mathcal{L}}_e$ denotes the Lagrangian for the electromagnetic field. 

Now, variation of the action~\eqref{1.1} with respect to the metric tensor component $g_{\mu\nu}$ yields the field equations in $f(R,\mathcal{T})$ gravity theory given by:
\begin{multline}\label{1.5}
G_{\mu\nu}=\frac{1}{f_R(R,\mathcal{T})}\left[8\pi T_{\mu\nu}+\frac{1}{2} f(R,\mathcal{T}) g_{\mu\nu}\right. \\ \left. -\frac{1}{2}R{f_R(R,\mathcal{T})}g_{\mu\nu} - (T_{\mu\nu}+\Theta_{\mu\nu}){f_T(R,\mathcal{T})} +8\pi E_{\mu\nu}\right],
\end{multline}
where we denote ${f_R (R,\mathcal{T})= \frac{\partial f(R,\mathcal{T})}{\partial R}}$, $\Theta_{\mu\nu}=\frac{g^{\alpha\beta}\delta T_{\alpha\beta}}{\delta g^{\mu\nu}}$ and ${f_\mathcal{T}(R,\mathcal{T})=\frac{\partial f(R,\mathcal{T})}{\partial \mathcal{T}}}$. Here ${\Box \equiv\partial_{\mu}(\sqrt{-g} g^{\mu\nu} \partial_{\nu})/\sqrt{-g}}$ is the D'Alambert operator, $R_{\mu\nu}$ is the Ricci tensor, $\nabla_\mu$ represents the covariant derivative associated with the Levi-Civita connection of $g_{\mu\nu}$, $G_{\mu\nu}$ is the Einstein tensor and $E_{\mu\nu}$ is the electromagnetic energy-momentum tensor. 

We define $T_{\mu\nu}$ and $E_{\mu\nu}$ as follows:
\begin{eqnarray}\label{1.6}
T_{\mu\nu}&=&\left( \rho+p \right)u_{\mu} u_{\nu}+p g_{\mu\nu}, \\ \label{1.7}
E_{\mu\nu}&=&\frac{1}{4\pi}\left(F^{\gamma}_{\mu} F_{\nu \gamma}-\frac{1}{4} g_{\mu \nu} F_{\gamma \beta} F^{\gamma \beta} \right),
\end{eqnarray}
where $u_{\mu}$ is the four velocity which satisfies the conditions $u_{\mu}u^{\mu}=1$ and $u^{\mu} {\nabla}_{\nu} u_{\mu}=0$, respectively, $\rho$ and $p$ represent matter density and pressure, respectively. In the present work, we consider $\mathcal{L}_m=-p$ and we obtain $\Theta_{\mu\nu}=-2 T_{\mu\nu} - p g_{\mu\nu}$. 

Now, the covariant divergence of Eq.~\eqref{1.5} reads
\begin{multline}\label{eqs}
{\nabla}^{\mu} T_{\mu\nu}=\frac{f_T \left(R,\mathcal{T}\right)}{8\pi-f_T \left(R,\mathcal{T}\right)} \left[\left(T_{\mu\nu}+\Theta_{\mu\nu} \right){\nabla}^{\mu} lnf_T \left(R,\mathcal{T}\right)\right. \\ \left.
+{{\nabla}^{\mu}}\Theta_{\mu\nu}-\frac{1}{2}g_{\mu\nu}{\nabla}^{\mu} T -\frac{8\pi}{f_T \left(R,\mathcal{T}\right)}{\nabla}^{\mu} E_{\mu\nu}\right].
\end{multline}

Now, if we consider the simplest linear form of the function $f(R, \mathcal{T})$ as $f(R, \mathcal{T})=R+2\chi \mathcal{T}$, where $\chi$ is the matter-geometry coupling constant, the field equation for $f(R, \mathcal{T})$ gravity theory reads
\begin{eqnarray}\label{1.9}
& G_{\mu\nu} = (8\pi+2\chi)T_{\mu\nu}+2\chi p g_{\mu\nu}+\nonumber \\ &\chi \mathcal{T} g_{\mu\nu}+8\pi E_{\mu\nu}
 = 8\pi \left(T^{eff}_{\mu\nu}+E_{\mu\nu}\right)=8\pi T_{ab},
\end{eqnarray}
where $T_{ab}=T^{eff}_{\mu\nu}+E_{\mu\nu}$ represents the energy-momentum tensor of the charged effective matter distribution and $T^{eff}_{\mu\nu}$ represents energy-momentum tensor of the effective fluid, i.e., ``normal'' matter and the new kind of fluid which originates due to the matter geometry coupling, given as
\begin{eqnarray} \label{1.10}
T^{eff}_{\mu\nu}=T_{\mu\nu}\left(1+\frac{\chi}{4\pi}\right)+\frac{\chi}{8\pi}\left(\mathcal{T}+2p\right) g_{\mu\nu}.
\end{eqnarray}

Substituting $f(R, \mathcal{T})=R+2\chi \mathcal{T}$ in Eq.~\eqref{eqs}, we obtain
\begin{multline} \label{1.11}
\left(4\pi+\chi \right) {\nabla}^{\mu} T_{\mu\nu}= \\-\frac{1}{2} \chi \Big[g_{\mu\nu} {\nabla}^{\mu}\mathcal{T}+2 {\nabla}^{\mu}\left(p g_{\mu\nu} \right)+\frac{8\pi}{\chi} E_{\mu\nu} \Big].
\end{multline}


\begin{figure*}[!htpb]
\centering
\subfloat[Effective pressure profile.\label{pressure}]{\includegraphics[width=0.44\textwidth]{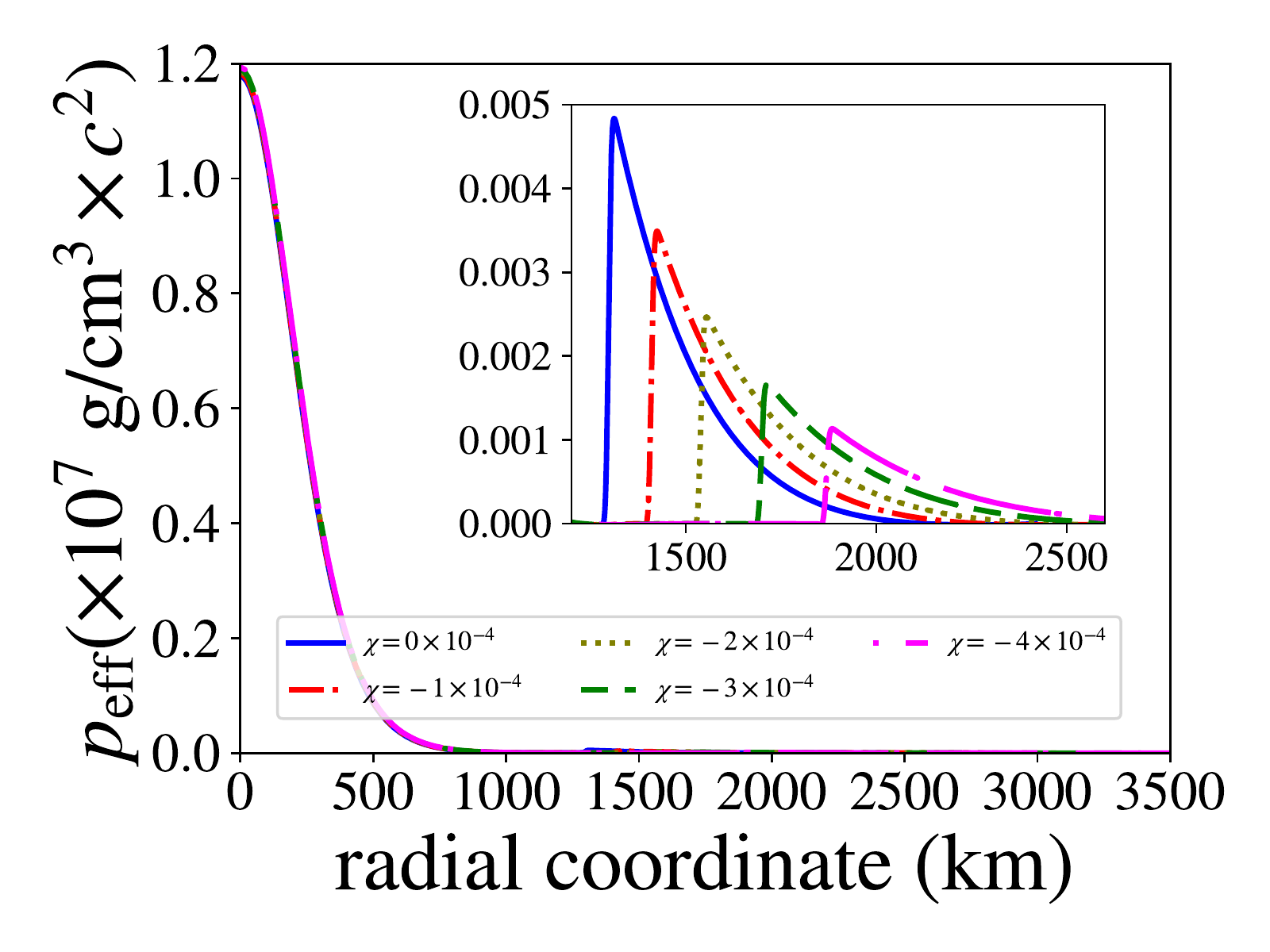}}\hfill
\subfloat[Effective energy density profile.\label{mass}]{\includegraphics[width=0.4\textwidth]{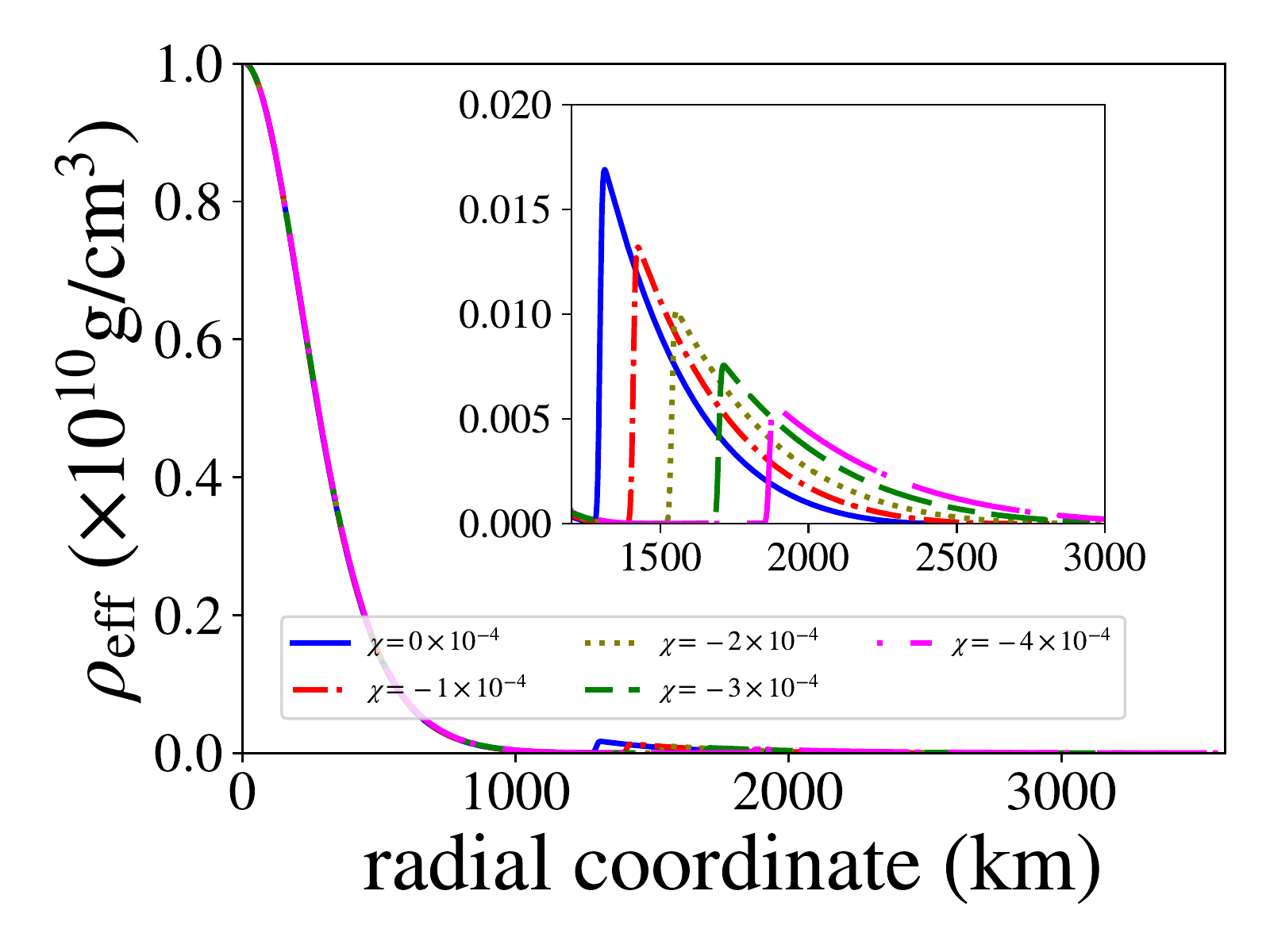}}\hfill
\subfloat[Charge profile.\label{charge}]{\includegraphics[width=0.4\textwidth]{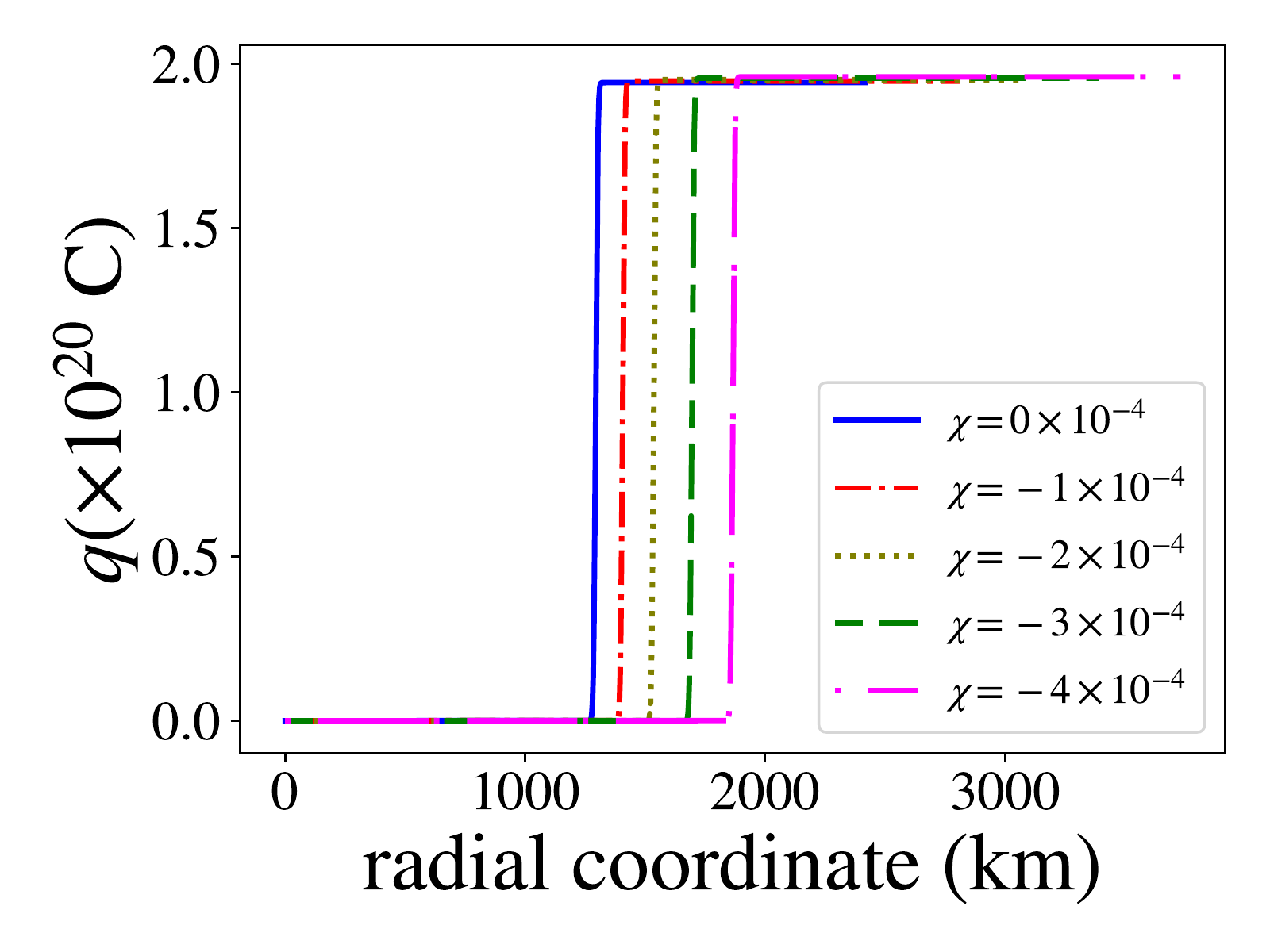}}\hfill
\subfloat[Electric field profile.\label{efield}]{\includegraphics[width=0.4\textwidth]{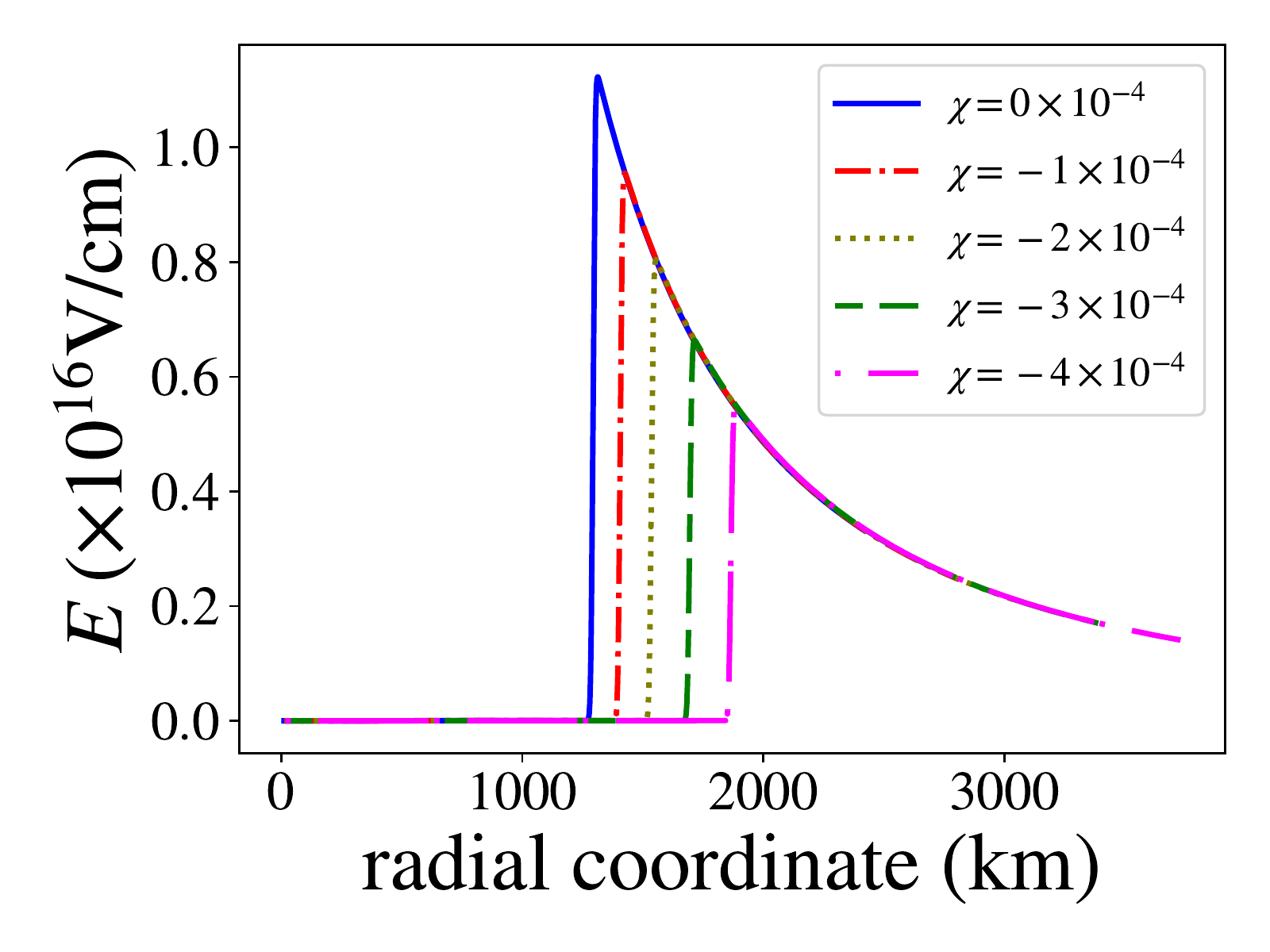}}
\caption{(color online) Profiles for several values of $\chi$, $\sigma=2\times 10^{20}$C and central density of $\rho_C=10^{10}{\rm g/cm^3}$ .} \label{general}
\end{figure*}



\begin{figure*}[!htpb]
\centering
\subfloat[Mass-radius relation of white dwarfs with varying $\chi$.\label{mass-radius1}]{\includegraphics[scale=0.52]{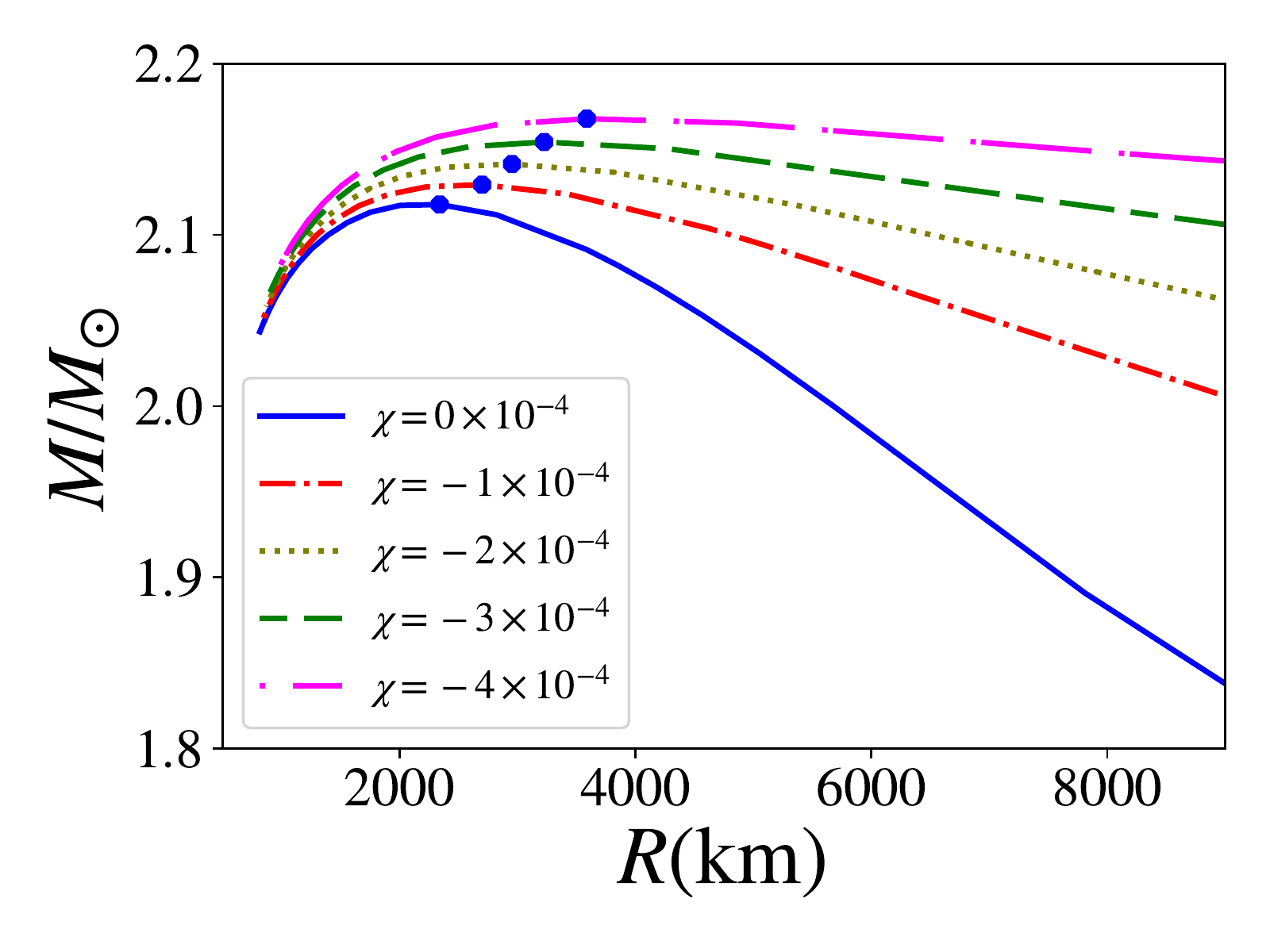}}\hfill
\subfloat[Mass-radius relation of white dwarfs with varying $\sigma$.\label{mass-radius2}]{\includegraphics[scale=0.52]{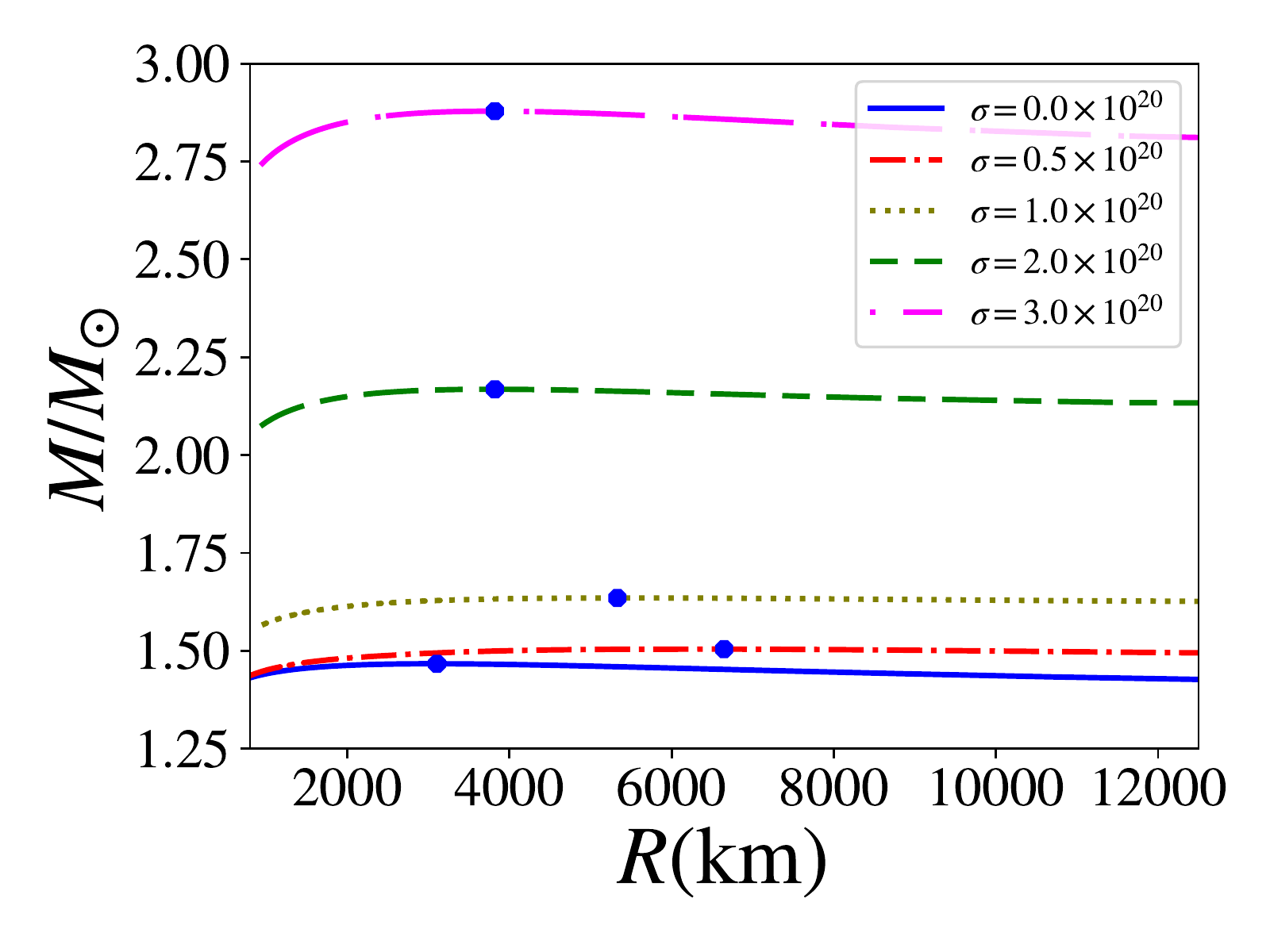}}
\caption{(color online) Mass-radius relation of white dwarfs for i) several values of $\chi$ and $\sigma=2\times 10^{20}$C and ii) several values of $\sigma$ and $\chi=-4 \times {10}^{-4}$.} \label{mass-radius}
\end{figure*}


\subsection{Stellar Equilibrium Equations}
Let consider the interior space-time is described by the metric as follows:
\begin{eqnarray}\label{1.13}
ds^2=e^{\nu(r)}dt^2-e^{\lambda(r)}dr^2-r^2(d\theta^2+\sin^2\theta d\phi^2),
\end{eqnarray} 
where the metric potentials $\nu$ and $\lambda$ are the function of the radial coordinate $r$ only.

Now substituting Eqs. \eqref{1.6} and \eqref{1.7} into Eq. \eqref{1.9} we find the explicit form of the Einstein field equation for the interior metric \eqref{1.13} as follows:
\begin{eqnarray}\label{1.21}
{{\rm e}^{-\lambda }} \left( {\frac {\lambda^{{\prime}}}{r}}-\frac{1}{r^2}\right) +\frac{1}{r^2}&=& \left(8\pi+3\chi\right)\rho-\chi p+\frac{q^2}{r^4} \nonumber \\ &=& 8\pi \rho^{\textit{eff}}+\frac{q^2}{r^4}, \\ \label{1.22}
{{\rm e}^{-\lambda}} \left( {\frac {\nu^{{\prime}}}{r}}+\frac{1}{r^2}\right) -\frac{1}{r^2}&=&\left(8\pi+3\chi\right)p-\chi\rho-\frac{q^2}{r^4}  \nonumber \\ &=& 8\pi  p^{\textit{eff}}-\frac{q^2}{r^4}, 
\end{eqnarray}
where `$\prime$' denotes differentiation with respect to the radial coordinate $r$. Here $\rho^{\textit{eff}}$ and $p^{\textit{eff}}$ represent effective density and pressure of the effective matter distribution, respectively, and are given by
\begin{eqnarray}\label{1.23}
& \rho^{\textit{eff}}=\rho+\frac{\chi}{8\pi}\left(3\rho-p\right), \\ \label{1.23a}
& p^{\textit{eff}}=p-\frac{\chi}{8\pi}\left(\rho-3p\right).
\end{eqnarray}

The further essential stellar structure equations required to describe static and charged spherically symmetric sphere in $f(R, \mathcal{T})$ gravity theory are given as
\begin{eqnarray}\label{1.24}
\frac{dm}{dr}&=&4\pi \rho r^2+\frac{q}{r} \frac{dq}{dr}+\frac{\chi}{2}\left(3\rho-p\right)r^2, \\ \label{1.25}
\frac{dp}{dr}&=&\frac{1}{\left[1+\frac{\chi}{8\pi+2\chi}\left(1-\frac{d\rho}{dp}\right)\right]}\Bigg\lbrace -\left(\rho+p\right)\bigg[\Big\lbrace 4\pi\rho r+\frac{m}{r^2} \nonumber \\ &-&\frac{q^2}{r^3}  -\frac{\chi}{2}\left(\rho-3p\right)r \Big\rbrace \bigg/ \left(1-\frac{2m}{r}+\frac{q^2}{r^2}\right)\bigg] \nonumber \\ &+& \frac{8\pi}{8\pi+2\chi}\frac{q}{4\pi r^4}\frac{dq}{dr} \Bigg\rbrace,
\end{eqnarray}  
where the metric potential $e^{\lambda}$ have the usual Reisner-Nordstr{\"o}m form
\begin{equation}
e^{\lambda}=1-\frac{2m}{r}+\frac{q^2}{r^2}.
\end{equation}

Clearly, due to the absence of matter out side the surface of the stellar system $\chi=0$ which leads to describe the exterior space-time by the exterior Reissner-Nordstr{\"o}m metric same as GR, reads
\begin{multline}\label{1.27}
ds^2 = \left(1 - \frac{2M}{r} +\frac{Q^2}{r^2}\right) dt^2- \frac{1}{\left(1 - \frac{2M}{r} + \frac{Q^2}{r^2}\right)} dr^2 \\ - r^2(d\theta^2 + sin^2\theta d\phi^2).
\end{multline}

In the present case, the modified TOV equation reads:
\begin{multline}\label{1.28}
 -p^{\prime}-\frac{1}{2}\nu^\prime \left(\rho+p\right)+\frac{\chi}{8\pi+2\chi}\left(\rho^{\prime}-p^{\prime}\right)\\ +\frac{8\pi}{8\pi+2\chi} \frac{q q^\prime}{4\pi r^4} =0,
\end{multline}

\section{Stellar properties}\label{secIII}
\subsection{Equation of State}

It is considered that the pressure and the energy density of the fluid contained in the spherical object are as follows
\begin{equation}\label{pressureEOS}
p\left(k_{F}\right)=\frac{1}{3 \pi^{2} \hbar^{3}} \int_{0}^{k_{F}} \frac{k^{4}}{\sqrt{k^{2}+m_{e}^{2}}} d k,    
\end{equation}
\begin{equation}\label{densityEOS}
    \rho\left(k_{F}\right)=\frac{1}{\pi^{2} \hbar^{3}} \int_{0}^{k_{F}} \sqrt{k^{2}+m_{e}^{2}} k^{2} d k+\frac{m_{N} \mu_{e}}{3 \pi^{2} \hbar^{3}} k_{F}^{3},
\end{equation}
where $m_e$ represents the electron mass, $m_N$ the nucleon mass, $\hbar$ is the reduced Planck constant, $\mu_e$ is the ratio between the nucleon number and atomic number for ions and $k_F$ represents the Fermi momentum of the electron. Equation \eqref{pressureEOS} states the electric degeneracy pressure and \eqref{densityEOS} gives the total energy density as the sum of the relativistic electron energy density (first term of the right-hand side) and the energy density related to the rest mass of nucleons (second term of the right-hand side). 

For numerical purposes we rewrite Eqs. \eqref{pressureEOS} and \eqref{densityEOS} as 
\begin{equation}
    p(x)=\epsilon_0 f(x),
\end{equation}
\begin{equation}
    \rho (x) = \epsilon_0 g(x),
\end{equation}
where
\begin{equation}
    f(x)=\frac{1}{24}\left[\left(2 x^{3}-3 x\right) \sqrt{x^{2}+1}+3 \operatorname{asinh} x\right],
\end{equation}
\begin{equation}
    g(x)=\frac{1}{8}\left[\left(2 x^{3}+x\right) \sqrt{x^{2}+1}-\operatorname{asinh} x\right]+1215.26 x^{3},
\end{equation}
with $\epsilon_0= m_e/\pi^2 \lambda_e^3$ and $x=k_F/m_e$ is the dimensionless Fermi momentum, $\lambda_e$ represents the electron Compton wavelength. In the above equation we take $\mu_e=2$.


\begin{table*}[t]
  \begin{center}
    \caption{The physical parameters of the charged white dwarfs in $f(R,\mathcal{T})$ gravity due to parametric values of $\chi$ and $\sigma=2\times 10^{20}$C.}
    \label{table1} 
    \begin{tabular}{cccccc} 
        \hline
      $\chi$ & $M/M_{\odot}$ & R (km) & $\rho_{C}^{\rm eff}{\rm (g/cm^3)}$ & $Q$(C)& $E {\rm(V/m)}$\\
        \hline
       $-0\times10^{-4}$ & 2.11 & 2201 & $1.45\times 10^{10}$ & $1.94\times 10^{20}$ & $3.59\times 10^{17}$\\
       $-1\times10^{-4}$& 2.13 & 2565 & $1.34\times 10^{10}$ & $1.94\times 10^{20}$ & $2.65\times 10^{17}$\\
       $-2\times10^{-4}$& 2.14 & 2954 &$1.23\times 10^{10}$  & $1.95\times 10^{20}$ & $2.01\times 10^{17}$\\
       $-3\times10^{-4}$& 2.15 & 3227 & $1.14\times 10^{10}$ & $1.95\times 10^{20}$ & $1.69\times 10^{17}$\\
       $-4\times10^{-4}$& 2.17 & 3820 & $9.60\times 10^{9}$ & $1.96\times 10^{20}$ & $1.21\times 10^{17}$ \\
       \hline
    \end{tabular}
  \end{center}
\end{table*}



\begin{table*}
  \begin{center}
    \caption{The physical parameters of the charged white dwarfs in $f(R,\mathcal{T})$ gravity due to parametric values of $\sigma$ and $\chi=-4\times 10^{-4}$.}
    \label{table2} 
    \begin{tabular}{cccccc} 
        \hline
      $\sigma$ (C) & $M/M_{\odot}$ & R (km) & $\rho_{C}^{\rm eff}{\rm (g/cm^3)}$ & $Q$(C)& $E {\rm(V/m)}$\\
        \hline
       $0.0\times10^{20}$  & 1.47  & 2940 & $3.37\times 10^{9}$  & - & -\\
       $0.5\times10^{20}$  & 1.50  & 6647 & $2.60\times 10^{9}$  & $4.94\times 10^{19}$ & $1.00\times 10^{16}$\\
       $1.0\times10^{20}$  & 1.63  & 5330 & $4.29\times 10^{9}$  & $9.86\times 10^{19}$ & $3.12\times 10^{16}$\\
       $2.0\times10^{20}$ & 2.17 & 3820  & $9.60\times 10^{9}$  & $1.96\times 10^{20}$ & $1.21\times 10^{17}$\\
       $3.0\times10^{20}$  & 2.88  & 3770 & $9.94\times 10^{9}$  & $2.94\times 10^{20}$ & $1.86\times 10^{17}$\\
       \hline
    \end{tabular}
  \end{center}
\end{table*}


\subsection{Electric Charge Profile}

We assume as in previous works that the star is mainly composed of degenerate material, so any charge present in the white dwarf would be concentrated close to the star's surface. Thus, following \cite{Carvalho2018,Negreiros2009Oct} we model the electric charge distribution in terms of a Gaussian distribution
\begin{equation}
    \rho_{e}=k \exp \left[-\frac{(r-R)^{2}}{b^{2}}\right],
\end{equation}
where $R$ is the radius of the star in the uncharged case, $b$ is the width of the electric charge distribution, in which we consider $b=10$~km. For small width of this layer the white dwarf structure does not change significantly.

Taking into consideration an arbitrary constant $\sigma$ given by
\begin{equation}
    \sigma=\int_{0}^{\infty} 4 \pi r^{2} \rho_{e} d r,
\end{equation}
we can estimate the proportionality constant $k$. Note that $\sigma$ would be the total charge with we were working in a flat space-time. Considering $\sigma$ as a comparison parameter we can estimate $k$ as
\begin{equation}
    8 \pi k=\sigma\left(\frac{\sqrt{\pi} b R^{2}}{2}+\frac{\sqrt{\pi} b^{3}}{4}\right)^{-1}.
\end{equation}

\section{Results}\label{secIV}

The mass of the charged WDs as a function of their total radii is shown in Fig.\ref{mass-radius} for five different values of $\chi$ and for $\sigma=2\times10^{20}$C. It is worth to cite that $\chi=0$ recovers GR results. 

In order to observe the electric charge distribution in the star, the effective pressure inside the WD as a function of the radial coordinate is showed in Fig.\ref{pressure}, where few values of $\chi$ and $\rho_C=10^{10}{\rm g/cm^3}$ are considered. In the figure, we can note that the pressure decays monotonically toward the baryonic surface, when it is attained, the pressure grows abruptly due to the beginning of the electrostatic layer. After this point, the pressure decrease with the radial coordinate until it attains the surface of the stars, which results in an electric charge distribution as a spherical shell close to the surface of the WD. For Fig.\ref{general} we took into account $\rho_C=10^{10}{\rm g/cm^3}$ and different values of $\chi$. Fig.\ref{mass} shows the effective energy density is as a function of the radial coordinate. 

In the Fig.~\ref{efield} the behavior of the electric field in the star is presented. We can note in the figure that the electric field exhibits a very abrupt increase from zero to $10^{16-17}{\rm V/m}$, this indicates that the baryonic surface ends and starts the electrostatic layer. The same behavior can be observed in Fig.\ref{charge} - interface between baryonic and electrostatic layers - where we present the charge profile.

As we can see in Fig.~\ref{mass-radius1} the mass of the stars grows as the total radius decreases until it attains a maximum mass point. It is important to remark that the maximum mass grows with the decrement of $\chi$. The total radius increases when fixed star masses are considered, which implies that the effects of the $f(R,\mathcal{T})$ gravity are very important in the determination of the stellar radius. Further, in Fig.~\ref{mass-radius1} for a fixed value of $\chi=-4\times {10}^{-4}$ the maximum mass of the white dwarfs increase as the values of $\sigma$ increase. Here in this work, we consider $\sigma=2\times10^{20}C$. This value of total charge has shown to saturate the electric field limit at the surface of the star, i.e., the Schwinger limit ($\sim1.3\times{10}^{18}$~V/m) for a mass of $2.199~M_{\odot}$ \cite{Carvalho2018}. We also can see in Fig.~\ref{mass-radius1} that the mass-radius curves tend to a plateau when $\chi$ is $\approx -4\times10^{-4}$. Evidently, the curves in Fig.~\ref{mass-radius} and obtained results feature a similar behavior in comparison with the mass-radius relations of the white dwarfs as reported by Carvalho and collaborators in Ref. \cite{Carvalho2017}.


\begin{figure}[!htpb]\centering
  \centering
  \includegraphics[scale=0.52]{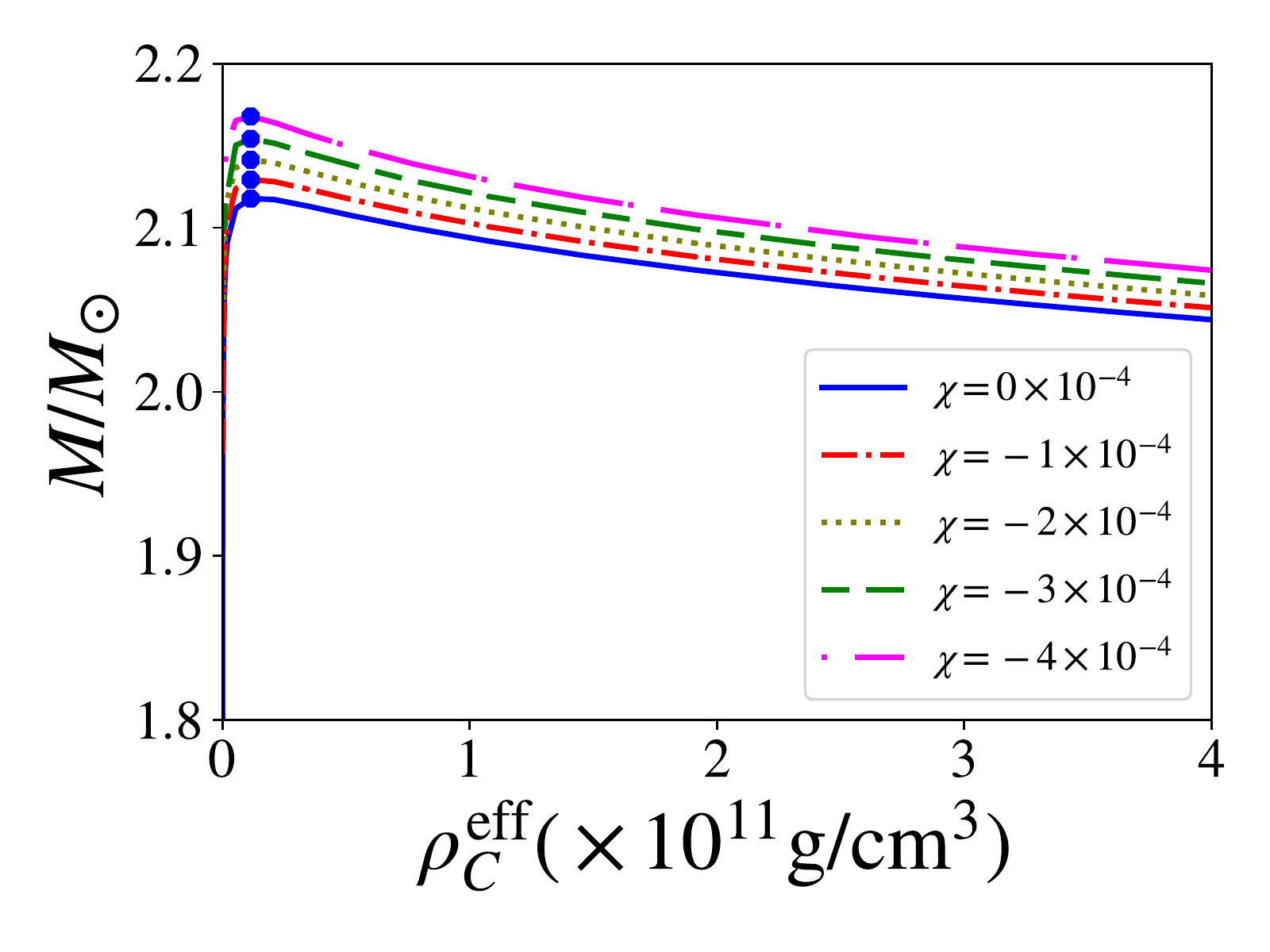}
  \caption{Mass-central density relation of white dwarfs for several values of $\chi$ and $\sigma=2\times 10^{20}$C.}\label{mass-density}
\end{figure}



\begin{figure}[!htpb]\centering
  \centering
  \includegraphics[scale=0.52]{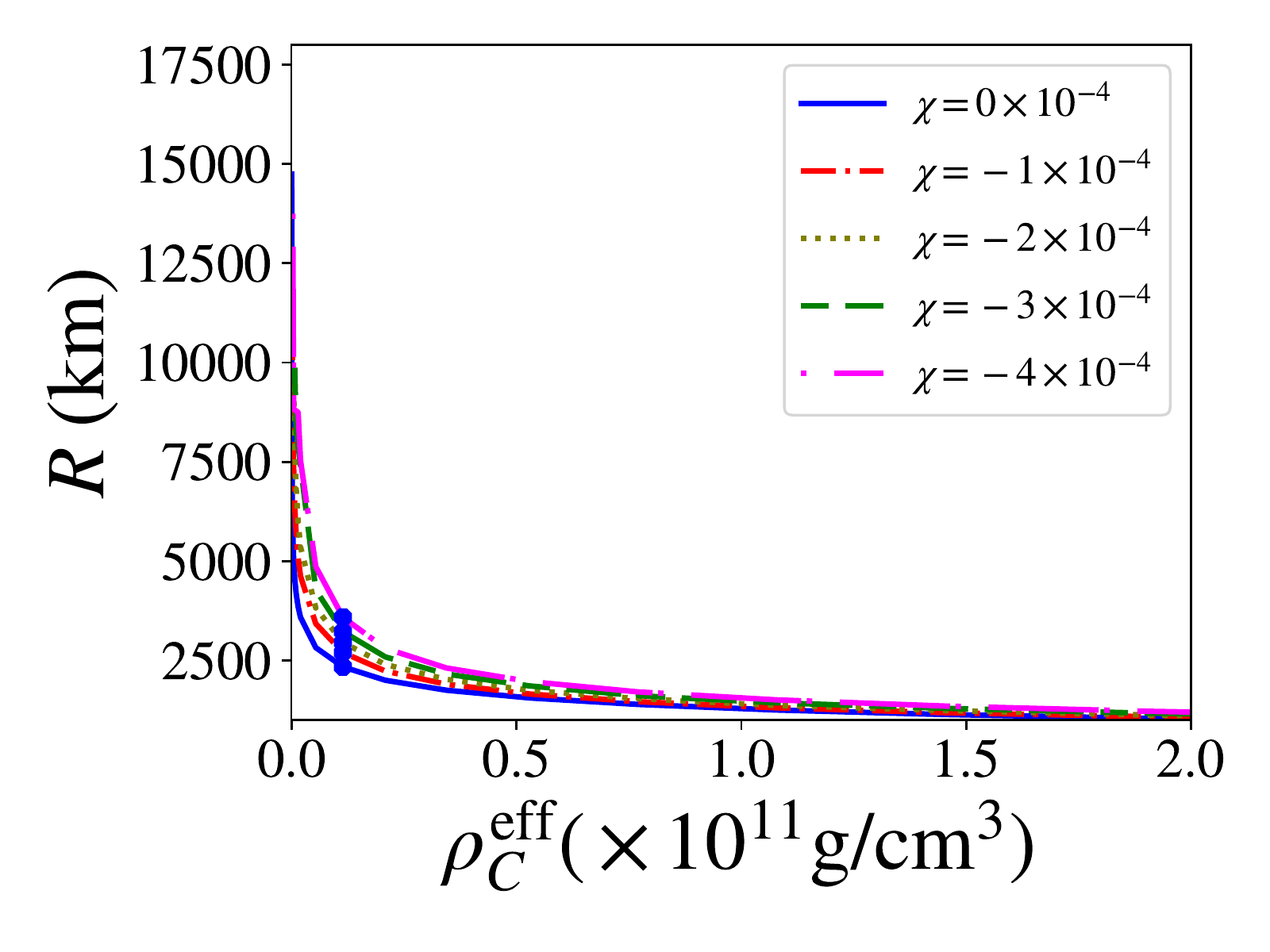}
  \caption{Central energy density versus total radius of white dwarfs for several values of $\chi$ and $\sigma=2\times 10^{20}$C.}\label{density-radius}
\end{figure}


In Fig. \ref{mass-density} we present the mass-central density relation of static, charged and non-charged WDs for five different values of $\chi$ and $\sigma=2\times 10^{20}$C. As in previous works \cite{Negreiros2009Oct,Liu2014,Arbanil2015,Carvalho2018,Deb2018epjc,Deb2018Dec} we can see that the charge produces a force, repulsive in nature, which helps the one generated by the radial pressure to support more mass against the gravitational collapse, so the masses in the charged case can be larger than in the non-charged one. We present also the radius-central density relation in Fig.\ref{density-radius}. To construct figures \ref{mass-density} and \ref{density-radius}, we used effective central energy density, defined as in Eq. \eqref{1.23}.


\begin{figure}[!htpb]\centering
  \centering
  \includegraphics[scale=0.52]{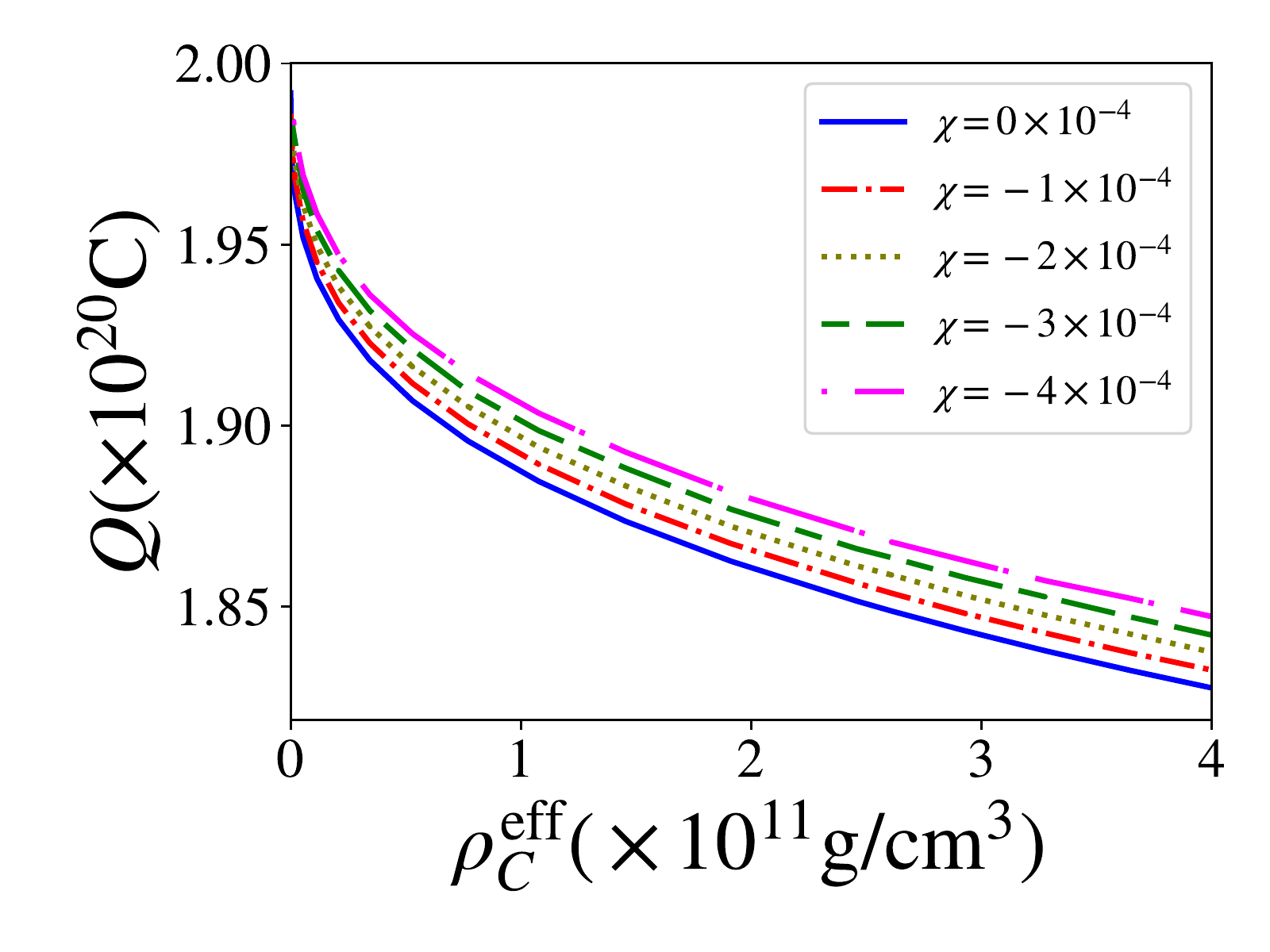}
  \caption{Total charge versus central energy density of white dwarfs for several values of $\chi$ and $\sigma=2\times 10^{20}$C.}\label{charge-density}
\end{figure}


In Fig.\ref{charge-density} is showed the total charge of the star as a function of the central effective energy density for the chosen parametric values of $\chi$ and $\sigma=2\times10^{20}$C. One can see that the total charge slightly varies with the increasing effective central density. 

In Table \ref{table1} we present the maximum masses for the charged WD in $f(R, \mathcal{T})$ gravity with their total radii and effective central energy densities for each value of $\chi$ used in this work. It is possible to note that more massive and large charged WDs are found with the decrement of $\chi$. We also note an important effect caused by the $f(R,\mathcal{T})$ gravity theory that is the increase of the radii, which contributes to the stability of the star, since it reduces the surface electric field. From table \ref{table1} one can realize that as the values of $\chi$ decrease the stellar system become more massive and larger in size turning itself into a less dense compact stellar object as predicted by Carvalho et al. in their study~\cite{Carvalho2018}. We have also predicted in Table \ref{table2} different physical parameters of the compact stellar system due to the variation of $\sigma$ for a chosen parametric value of $\chi=-4\times {10}^{-4}$. Table \ref{table2} features that with the increasing values of $\sigma$ as usually the mass of the white dwarfs increase along with their surface charge and electric field, whereas the stellar system becomes gradually denser as it's central density increase gradually with the increasing values of $\sigma$. We also find for $\sigma$ in the range $2\times {10}^{20}-3 \times {10}^{20}$~C the present $f(R,\mathcal{T})$ model is suitable to predict different physical parameters of the highly super-Chandrasekhar white dwarfs having mass $2.17-2.88~M_\odot$.

\section{Conclusions}\label{secV}

In this article, we investigate the effects of a specific modified theory of gravity, namely, the $f(R,\mathcal{T})$ gravity, in the structure of charged white dwarfs. The procedure started from the derivation of the hydrostatic equilibrium equation for such a theory, with the addition of the charged effects. We suppose a Gaussian ansatz for the net charge distribution. 

The main goal was to check the imprints of the extra material terms that come from the $\mathcal{T}$-dependence of the theory on charged WD properties.

The equilibrium configurations of charged white dwarfs were analyzed for $f(R, \mathcal{T})=R+2\chi \mathcal{T}$ with different values of $\chi$ and central densities. We observed that the charged white dwarfs can be affected by the extended theory of gravity in the maximum mass and radius depending on the value of $\chi$.

We found that for $\chi=-4\times10^{-4}$ and $\sigma=3 \times {10}^{20}$~C, the maximum mass of the charged WD is $2.88~M_\odot$, and the radii have considerable increasing. This larger radius yields a smaller surface electric field, thus enhancing the stellar stability of charged stars.

The possibility of explaining the highly super-Chandrasekhar limiting mass white dwarfs as progenitors of the peculiar over-luminous super-SNeIa was raised by Deb and collaborators in their work~\cite{Deb2018Dec}. In the present work, we have successfully explained the highly super-Chandrasekhar limiting mass white dwarfs having mass $2.17-2.88~M_\odot$ which remained hardly explained in the framework of GR. Importantly, the present work not only pushing the maximum mass limit for white dwarfs beyond the standard value of the Chandrasekhar mass-limit but also plausibly explaining the requirement of the application of $f(R, \mathcal{T})$ gravity theory in studying astrophysical observations.

\section{acknowledgments}
FR would like to thank CAPES (Coordena\c{c}\~ao de Aperfei\c{c}oamento Pessoal de N\'ivel Superior) for financial support. GAC thanks CAPES grants PDSE/88881.188302/2018-01 and PNPD/88887.368365/2019-00. A part of this work was completed by DD when he was visiting the IUCAA, and DD gratefully acknowledges warm hospitality there. MM acknowledge CAPES, CNPq and FAPESP thematic project $13/26258-4$. All computations were performed in open source software and the authors are sincerely thankful to open source community \cite{Python,Octave,wxmaxima}

\end{document}